\documentclass[%
 reprint,
 amsmath,amssymb,
 aps,
]{revtex4-2}
\usepackage{standalone}
\usepackage{xcolor}
\usepackage{graphicx}
\usepackage{dcolumn}
\usepackage{bm}
\usepackage{pgfplots}
\pgfplotsset{compat = newest}
\usepackage{pgfplotstable}
\usetikzlibrary{positioning, decorations.pathreplacing, calc, intersections, pgfplots.groupplots,fadings}
\usetikzlibrary{arrows,intersections}
\usepackage{tikz-cd}
\usepackage{xcolor}
\usetikzlibrary{fadings}
\usetikzlibrary{decorations.markings}
\pgfkeys{
  /pgf/arrow keys/.cd,
  pitch/.code={%
    \pgfmathsetmacro\pgfarrowpitch{#1}
    \pgfmathsetmacro\pgfarrowsinpitch{abs(sin(\pgfarrowpitch))}
    \pgfmathsetmacro\pgfarrowcospitch{abs(cos(\pgfarrowpitch))}
  },
}

\pgfdeclarearrow{
  name = Cone,
  defaults = {       
    length     = +2.3pt +3.4,
    width'     = +0pt +0.2,
    line width = +0pt 1 1,
    pitch      = +0, 
  },
  cache = false,     
  setup code = {},   
  drawing code = {   
    \pgfmathsetmacro\pgfarrowhalfwidth{.38\pgfarrowwidth}
    \pgfmathsetmacro\pgfarrowhalfwidthsin{\pgfarrowhalfwidth*\pgfarrowsinpitch}
    \pgfpathellipse{\pgfpointorigin}{\pgfqpoint{\pgfarrowhalfwidthsin pt}{0pt}}{\pgfqpoint{0pt}{\pgfarrowhalfwidth pt}}
    \pgfusepath{fill}
    \pgfmathsetmacro\pgfarrowlengthcos{\pgfarrowlength*\pgfarrowcospitch}
    \pgfmathparse{\pgfarrowlengthcos>\pgfarrowhalfwidthsin}
    \ifnum\pgfmathresult=1
      \pgfmathsetmacro\pgfarrowlengthtemp{\pgfarrowhalfwidthsin*\pgfarrowhalfwidthsin/\pgfarrowlengthcos}
      \pgfmathsetmacro\pgfarrowwidthtemp{\pgfarrowhalfwidth/\pgfarrowlengthcos*sqrt(\pgfarrowlengthcos*\pgfarrowlengthcos-\pgfarrowhalfwidthsin*\pgfarrowhalfwidthsin)}
      \pgfpathmoveto{\pgfqpoint{\pgfarrowlengthcos pt}{0pt}}
      \pgfpathlineto{\pgfqpoint{\pgfarrowlengthtemp pt}{ \pgfarrowwidthtemp pt}}
      \pgfpathlineto{\pgfqpoint{\pgfarrowlengthtemp pt}{-\pgfarrowwidthtemp pt}}
      \pgfpathclose
      \pgfusepath{fill}
    \fi
    \pgfpathmoveto{\pgfpointorigin}
  }
}

\immediate\write18{pdflatex myletter}

\usepackage{graphicx}
\usepackage{amsmath}
\usepackage{xr}
\usepackage{pifont}
\usepackage{amssymb}
\usepackage{comment}
\usetikzlibrary{plotmarks}
\usepackage{float}
\usepackage{xcolor}
\usepackage{mathtools}
\usepackage{tikz}
\usepackage{physics}
\usepackage{mathtools}
\usepackage{enumitem}%
\usepackage{pgfplots}
\usepackage{pgfplotstable}
\usepackage{epsfig}
\usepackage{calculus}
\usepackage{calc}
\usetikzlibrary{calc,angles,quotes}
\usepackage{pgfplots}
\pgfplotsset{compat=1.18} 
\usepgfplotslibrary{ternary}
\usetikzlibrary{shadows}

\definecolor{darkblue}{RGB}{8, 66, 102}
\definecolor{darkblue2}{RGB}{42, 102, 143}
\definecolor{darkblue3}{RGB}{186, 199, 216}
\definecolor{darkred}{RGB}{231, 117, 0}
\definecolor{bluegrey}{RGB}{42, 102, 143}

\tikzset{
    vertex/.style = {
        circle,
        fill      = darkblue,
        outer sep = 2pt,
        inner sep = 2pt,
    }
}
\tikzset{
    vertex1/.style = {
        circle,
        fill      = darkred,
        outer sep = 2pt,
        inner sep = 2pt,
    }
}
\usepackage{xr}
\makeatletter

\newcommand*{\addFileDependency}[1]{
\typeout{(#1)}
\@addtofilelist{#1}
\IfFileExists{#1}{}{\typeout{No file #1.}}
}\makeatother

\newcommand*{\myexternaldocument}[1]{%
\externaldocument{#1}%
\addFileDependency{#1.tex}%
\addFileDependency{#1.aux}%
}

\myexternaldocument{Suppl-thirdrevision}
\makeatletter
\newcommand{\specialnumber}[1]{%
  \def\tagform@##1{\maketag@@@{(\@@italiccorr#1\unskip\ignorespaces##1)}}%
}
\newcommand{\specialeqref}[2]{\begingroup
  \def\tagform@##1{\maketag@@@{(\@@italiccorr#2\unskip\ignorespaces##1)}}%
  \eqref{#1}\endgroup}
\makeatother

\begin{document}

\preprint{APS/123-QED}

\title{Emergence of Light Cones in Long-range Interacting Spin Chains Is Due to Destructive Interference
}%

\author{Peyman Azodi}
\email{pazodi@princeton.edu}
\author{Herschel A.Rabitz}%

\affiliation{%
 Department of Chemistry, Princeton University, Princeton, New Jersey 08544, USA 
}%




\date{\today}

\begin{abstract}

{
Despite extensive research on long-range interacting quantum systems, the physical mechanism responsible for the emergence of light cones remains unidentified. This work presents a novel perspective on the origins of locality and emergent light cones in quantum systems with long-range interactions. We identify a mechanism in such spin chains where effective entanglement light cones emerge due to destructive interference among quantum effects that entangle spins. Although long-range entangling effects reach beyond the identified light cone, due to destructive interference, entanglement remains exponentially suppressed in that region, ultimately leading to the formation of the light cone. We demonstrate that this interference not only drives but is also necessary for the emergence of light cones. Furthermore, our analysis reveals that reducing the interaction range weakens this interference, surprisingly increasing the speed of entanglement transport—an effect that opens new experimental opportunities for investigation.

}

\end{abstract}

\maketitle


\section{Introduction} 
Understanding the propagation of entanglement in many-body quantum systems is a cornerstone of modern quantum science, with critical implications for both fundamental research and practical applications  \cite{ langen2013local,doi:10.1126/science.aaf6725,lewis2019dynamics,gogolin2016equilibration, kinoshita2006quantum, rigol2008thermalization,srednicki1994chaos,deutsch1991quantum,nandkishore2017many, pal2010many, lukin2019probing,bernien2017probing}. Despite decades of progress in mapping light cones and constraining information propagation velocities, one pivotal question remains unresolved: what physical mechanism underlie locality, i.e., the emergence of light cones, in quantum systems with long-range interactions? In this work, we reveal a previously unrecognized phenomenon—quantum destructive interference—as the driving force behind light cone formation.
\par In relativistic quantum field theory, the spread of correlations is strictly limited by a light cone \cite{peskin2018introduction}, governed by the speed of light. However, in non-relativistic quantum systems, the speed of light does not play such a role. Despite the latter circumstance, the Lieb-Robinson (LR) bound establishes an effective light cone structure in short-range interacting quantum systems, where correlations decay exponentially beyond this boundary \cite{lieb1972finite}. The word effective emphasizes the existence of tunneling effects beyond this boundary \cite{explanation5}. Over the past few decades, the LR theorem has been refined \cite{hastings2004lieb,hastings2006spectral, junemann2013lieb, kuwahara2024effective, gong2023long, PhysRevLett.104.190401,kuwahara2020strictly, PhysRevA.101.022333, matsuta2017improving, burrell2007bounds,defenu2023long, chen2023speed} and validated \cite{cheneau2012light, doi:10.1126/science.1248402, jurcevic2014quasiparticle, PhysRevX.8.021070, Storch_2015} across a wide range of quantum systems, include Long-Range Interacting (LRI) systems. In this class of systems, due to the long-range (all-to-all) nature of interactions, entangling effects reach beyond any possible light cone structures, potentially leading to the absence of any light cones; However, various forms of light cones continue to persist in the LRI regime \cite{foss2015nearly,kuwahara2020strictly, defenu2023long, cevolani2018universal,frerot2018multispeed, buyskikh2016entanglement, chen2023speed, xu2024scrambling, chen2019finite, nachtergaele2010lieb, bonnes2014light, defenu2023long, chen2023speed}. 
\par The goal of this paper is not to refine existing LR bounds or validate established knowledge. Instead, our objective is to uncover the physical phenomenon that underpins the emergence of light cones, which, to the best of our knowledge, has been overlooked in the literature. Specifically, we demonstrate that quantum destructive interference drives the emergence of light cones in long-range interacting quantum systems, underscoring the role of quantum phases in the emergence of locality in quantum systems. This new insight provides not only a foundational understanding of the concept of locality in quantum systems but also a deeper perspective on the \textit{emergent} nature of relativistic principles within quantum dynamics—an area of ongoing significant effort \cite{carlip2014challenges}.

We show that destructive interference is both necessary and sufficient for the emergence of light cones in LRI quantum chains. This conclusion is substantiated by several key observations. First, we demonstrate that light cones arise explicitly as a consequence of destructive interference (sufficiency). Second, we show that in the absence of destructive interference, light cones lose a fundamental property—observed and well-documented in the literature (necessity)—namely, their structural independence from the interaction length in LRI spin chains \cite{tran2020hierarchy, PhysRevLett.124.180601}. Third, we establish that the breakdown of destructive interference coincides with the disappearance of light cones, aligning with prior theoretical and numerical findings. Finally, we present both analytical and numerical evidence of a counterintuitive phenomenon: the acceleration of entanglement propagation—and the corresponding diminishing of light cones—when the interaction range is shortened, which weakens destructive interference. This finding is particularly surprising, as one would intuitively expect shorter interaction ranges to slow down entanglement propagation. These results collectively highlight the central role of destructive interference in shaping the dynamics of long-range quantum systems, which has novel implication to quantum information processing \cite{chen2023speed}.

\section{Spin Chain Model}

\par We consider one-dimensional (1D) spin chains with the Hamiltonian  
\begin{equation}\label{hamil22}
    \mathbf{H}=\sum _{j< l}^N
    \mathbf{J}(|j-l|) H_{j,l},
\end{equation}
where $H_{j,l}$ (satisfying $||H_{j,l}||=1$) is a Hermitian operator acting on sites $j$ and $l$. The interaction function $\mathbf{J}(n)$ is assumed to be real, symmetric ($\mathbf{J}(n) = \mathbf{J}(-n)$), and absolutely convergent, i.e.,  
\begin{equation}
\sum_{n} |\mathbf{J}(n)|<\infty.
\end{equation}
A special case is the power-law decaying interaction function, given by  
\begin{equation}\label{PD}
  \mathbf{J}(n)=\mathbf{J}(|j-l|)=\frac{J}{|j-l|^p}, 
\end{equation}
where $J$ represents the interaction strength and $p$ determines the interaction length. For simplicity, we set $\mathbf{J}(0) = 0$ and $\mathbf{J}(1) = 1$, implying $J = 1$. Additionally, we set $\hbar = 1$ throughout our formulation. The absolute convergence of the interaction function requires $p > 1$.  

\par In this setting, an arbitrary local observable $O$ ($||O||=1$) evolves as  
\begin{equation}
O(t) = U^{\dag} (t) O  U (t),
\end{equation}  
where $U(t) = e^{-i\mathbf{H}t}$ is the unitary time evolution operator generated by the Hamiltonian (\ref{hamil22}). The Taylor expansion of $U(t)$,  
\begin{equation}
    U(t) = \sum_{r} \frac{(-it\mathbf{H})^r}{r!},
\end{equation}  
contains hopping terms embedded in $\mathbf{H}^r$, which are induced by the interaction function $\mathbf{J}(n)$. These terms are given by the repeated convolution of $\mathbf{J}(n)$, expressed as (see also Figure \ref{fig11}):  
\begin{equation}\label{expansionJt}
\begin{split}
        {{\mathbf{J}}^{*r}(q)} =&\sum _{\substack{d_1,d_2,\cdots, d_r \in \mathbb{Z}- \{0\}, \sum d_i=q}}\prod_i \mathbf{J}(d_i).
\end{split}
\end{equation}
Here, $q$ denotes the total displacement along the chain resulting from $r$ successive hopping events, with individual hopping distances given by $d_1, \dots, d_r$.  

\par Motivated by this observation, and considering the presence of both $U$ and $U^\dag$ in the expression for $O(t)$, we study the evolution of the following quantity:  
\begin{multline}\label{tay0}
      \mathcal{Q}_{q}(t)=\sum_{r=0} {\alpha _r} t^{2r},
        \\ \alpha_r \equiv \frac{1}{(2r)!}\sum_{r_1+r_2=2r} \binom{2r}{r_1} (-1)^{r_1}\mathbf{J}^{*r_1}(q) \mathbf{J}^{*r_2}(q). 
\end{multline}
Since $||H_{j,l}|| = ||O|| = 1$, the function $\mathcal{Q}_{q}(t)$ serves as an upper bound for variations in the expectation value of $O$, evaluated with respect to an arbitrary wavefunction, when transported to distance $q$ under the unitary evolution dictated by $\mathbf{H}$ \cite{gong2014persistence, lieb1972finite}, while taking into account contributions from both $U$ and $U^\dag$ at distance $q$.  

\par Notably, in Section II of \cite{suppl}, we prove that $\mathcal{Q}_{q}(t)$ exactly quantifies entanglement production at distance $q$, as a result of a local quench, and in the free-particle limit. Hence, in the remainder of this paper, we will refer to $\mathcal{Q}_{q}(t)$ as entanglement produced at distance $q$.
\par Furthermore, the series expansion in (\ref{tay0}) is restricted to even powers of $t$. This follows from the fact that $\mathbf{J}(n)$ is a real function, which ensures that we can always take $ \mathcal{Q}_{q}(t) = \mathcal{Q}_{q}(-t)$, as discussed in \cite{azodi2023exact}. 

\section{Characterizing Light Cones} \label{lightchar}\par In the short-range limit, i.e., {Nearest-Neighbor (NN) interactions}, we have $\alpha_r = 0$ for $r < q$, and the first nonzero term in the Taylor series expansion (\ref{tay0}) is given by \cite{azodi2023exact, azodi2023dynamics}:
\begin{equation}\label{first}
    \mathcal{Q}_{q}(t) \propto \left(\frac{t}{q}\right)^{2q}.
\end{equation}  
This term dominates the entanglement expansion for $t < q$ \cite{explanation3} and explicitly dictates a {linear light cone} structure for $\mathcal{Q}_{q}(t)$, as expected \cite{lieb1972finite}. (Note that since we set $\hbar = 1$ and $\mathbf{J}(1) = 1$, time $t$ and distance $q$ can be directly compared, e.g., $t \leq q$.)  

Thus, (\ref{first}), which corresponds to $r = q$ in the expansion (\ref{tay0}), governs {early entanglement growth} at distance $q$ (where \textit{early} refers to $t \leq q$). Due to the interplay between $r$, $q$, and $t$, we can {identify and define} the emergent \textit{linear light cone} by the term $r = q$ in the Taylor series (\ref{tay0}). Similarly, terms with $r < q$ ($r > q$) correspond to regions \textit{outside} (\textit{inside}) the light cone \cite{explanation1}.  

\par Analogously, when the coefficients $\alpha_r$ vanish (or are exponentially suppressed) for $r < \frac{q}{2}$, and the early entanglement dynamics ($t < \frac{q}{2}$) satisfy  
\begin{equation}
  \mathcal{Q}_{\mathcal{M}}(t) \propto \left( \frac{2t}{q} \right)^{q},
\end{equation}  
we can identify the emerging linear light cone with $2r \approx q$. The rationale for this characterization will be clarified in subsection \ref{sectheorem}. Moreover, the conditions $2r < q$ ($2r > q$) define the regions {outside} ({inside}) the light cone in this scenario. We will use this approach to rigorously {prove the existence of a light cone} in Section \ref{secIV}. This proof is established by demonstrating that $\alpha_r \approx 0$ for $2r < q$, despite the extensive hopping effects— a direct consequence of {destructive interference}.

\section{main result: Emergence of Light Cones via Destructive Interference}\label{secIV}
For long-range interacting systems, the hopping process in (\ref{expansionJt}) leads to an accumulation of entangling effects even on short time scales. This occurs due to the exponential increase in the number of possible hopping paths that can connect any two points on the chain for $r > 2$. In fact, for $r > 2$, there exist infinitely many $r$-step hopping pathways that connect any two sites (for infinitely long chains).  

Despite this accumulative feature, one might expect that the formation of light cones results from the insignificance of the accumulated effects, attributed to the attenuation of the interaction function at long distances. However, this cannot be the underlying reason, as it does not explain the observation that the structure of light cones is largely independent of the interaction range \cite{tran2020hierarchy, PhysRevLett.124.180601}. We will further discuss this issue in Subsection \ref{nec}.

In this Section, having ruled out the above attenuation-based explanation, we demonstrate that light cone formation is entirely governed by the cancellation of terms in the expansion (\ref{tay0}) for $\alpha_r$. Specifically, we prove that destructive interference is not only sufficient (as shown in Subsection \ref{sectheorem}) but also necessary (as shown in Subsection \ref{nec}) for the emergence of light cones.  

\subsection{Two Types of Interaction Functions}\label{twotypes}

Before proceeding to the main proof of the role of destructive interference, it is essential to distinguish between two types of interaction functions (Type 1 and Type 2), as illustrated in Figure \ref{fig11}. Destructive interference occurs specifically for Type 1 functions.

\begin{itemize} 
    \item \textbf{Type 1}: The dominant terms in the expansion (\ref{expansionJt}) arise when most $d_i$'s are of order $1$, with only a few $d_i$'s of order $q$. 
    \item \textbf{Type 2}: The dominant terms in expansion (\ref{expansionJt}) correspond to a nearly uniform distribution of $d_i$'s, where the $d_i$'s are of order $q/r$. 
\end{itemize}

In Section IV of \cite{suppl}, we prove that the classification of an interaction function $\mathbf{J}(n)$ is determined by the quantity $\mathbf{J}''- \mathbf{J}'\mathbf{J}$, where $'$ denotes differentiation with respect to $n$, assuming the function is continuous. We show that as this quantity increases, Type 1 terms in (\ref{expansionJt}) become more dominant. Furthermore, in Subsection III.D of \cite{suppl}, we demonstrate that for $p>2$, the power-law interaction (\ref{PD}) falls under Type 1.

In the next subsection, we prove the existence of destructive interference for power-law interactions with $p>2$. However, the proof can be readily extended to all Type 1 interaction functions.

\begin{figure}
    \centering
    \begin{tikzpicture}[>=stealth]

\begin{scope} [xscale=0.85]

  \coordinate (o) at (0.6,0);
  \coordinate (i) at (8.4,0);
 
         \draw node[vertex1] (a) at (1,0) {};
  \draw node[vertex] (b) at (2,0) {};
  \draw node[vertex] (c) at (3,0) {};
  \draw node[vertex] (d) at (4,0) {};
  \draw node[vertex] (e) at (5,0) {};
  \draw node[vertex] (f) at (6,0) {};
  \draw node[vertex] (g) at (7,0) {};
  \draw node[vertex] (h) at (8,0) {};
  \draw node[vertex] (i) at (9,0) {};
  \node[below, yshift=-2pt] at (i) {\footnotesize$q$};
 \node[below, yshift=-2pt] at (b) {\footnotesize$1$};

  \draw[>->,, thick, darkred] (a) .. controls (1.5,0.3) .. (b); \node at (1.5,0.5) {\footnotesize $\mathsf{d_1=1}$ };
  \draw[>->,, thick, darkred] (b) .. controls (2.5,0.3) .. (c);
  \draw[>->,, thick, darkred] (c) .. controls (3.5,0.3) .. (d);
  \draw[>->,, thick, darkred] (d) .. controls (5,0.45) and (8,0.45) .. (i);\node at (6.5,0.6) {\footnotesize $\mathsf{d_4=5}$ };


  \draw[-, bluegrey]  (a) -- (b) (b)--(c) (c) --(d) (d)--(e) (e)--(f) (f)--(g) (g)--(h) (h)--(i);
  \draw[->, bluegrey, thick] (i)-- +(0.5,0);
    \draw[->, bluegrey, thick] (a)-- +(-0.5,0);
    \node at (0.2,0.8) {\small \textsf{Type 1:} };
    \node at (-0.3,0) {\footnotesize $\mathsf{r=4}$: };

\end{scope}

\begin{scope}[yshift=-1.5cm, xscale=0.85]
       \draw node[vertex1] (a) at (1,0) {};
   
  \draw node[vertex] (b) at (2,0) {};
  \draw node[vertex] (c) at (3,0) {};
  \draw node[vertex] (d) at (4,0) {};
  \draw node[vertex] (e) at (5,0) {};
  \draw node[vertex] (f) at (6,0) {};
  \draw node[vertex] (g) at (7,0) {};
  \draw node[vertex] (h) at (8,0) {};
  \draw node[vertex] (i) at (9,0) {};
  \node[below, yshift=-2pt] at (i) {\footnotesize$q$};
 \node[below, yshift=-2pt] at (b) {\footnotesize$1$};

  \draw[>->,, thick, darkred] (a) .. controls (2,0.4) .. (c);
  \draw[>->,, thick, darkred] (c) .. controls (4,0.4) .. (e);
  \draw[>->,, thick, darkred] (e) .. controls  (6,0.4) .. (g);
  \draw[>->,, thick, darkred] (g) .. controls  (8,0.4) .. (i);

  \draw[-, bluegrey]  (a) -- (b) (b)--(c) (c) --(d) (d)--(e) (e)--(f) (f)--(g) (g)--(h) (h)--(i);
  \draw[->, bluegrey, thick] (i)-- +(0.5,0);
    \draw[->, bluegrey, thick] (a)-- +(-0.5,0);
    \node at (0.2,0.5) {\small \textsf{Type 2:} };
    \node at (-0.3,0) {\footnotesize $r=4$: };
    \end{scope}

\end{tikzpicture}
    \caption{Schematic illustration of the hopping process at distance $q$ in $r$ steps, given by the $r$-th order convolution of $\mathbf{J}(n)$, as described in (\ref{expansionJt}). The figure distinguishes two different types of hopping, as defined in \ref{twotypes}. Type 1 consists of hoppings of order $1$ accompanied by a long jump, while Type 2 corresponds to a more uniform distribution of hoppings.
}
    \label{fig11}
\end{figure}


\subsection{Mechanism of Destructive Interference}\label{mech}
Here we calculate $\alpha_r$ given by the series expansion (\ref{tay0}). For this purpose, we first simplify the involved high-order convolutions ($\mathbf{J}^{* r} (q)$) of the power-law decaying interaction (\ref{PD}). Given that for $p>2$, the power-law interaction function (\ref{PD}) is Type 1, the high-order convolution (defined in (\ref{expansionJt})) of this function can be expanded as
\begin{equation}\label{expansionJt1}
\begin{split}
    {{\mathbf{J}}^{*r}(q)} =&\frac{\binom{r}{1}}{(q-r+1)^p}+ \frac{\binom{r}{2}}{\big (2(q-r)\big )^p} + \cdots.    
\end{split}
\end{equation}
Here, the first term corresponds to $d_1=d_2=\cdots =d_{r-1}=1$ and $d_r=q-r+1$ in (\ref{expansionJt}). Since any of the $d_i$'s (among $d_1, d_2, \cdots, d_r)$ can be chosen to equal $q-r+1$, there are $\binom{r}{1}$ such terms, which gives the numerator of the first fraction in this expansion. The second term in (\ref{expansionJt1}) corresponds to $d_1=d_2\cdots=d_{r-2}=1$, $d_{r-1}=2$, and $d_{r}=q-r$ in the expansion (\ref{expansionJt}). Simple combinatorial arguments dictate that there exist $\binom{r}{2}$ such terms.

\par We now focus on the first term in the expansion (\ref{expansionJt1}). A similar set of derivations holds for any other Type $1$ term in this series, all of which have been taken into account in the full proof in Section III of \cite{suppl}. When $0<r<q$, the first term in (\ref{expansionJt1}) can be written as
\begin{equation}\label{tay}
\begin{split}
    \frac{\binom{r}{1}}{(q-r+1)^p}=&
    \frac{r}{q^p} \left(1-\frac{r-1}{q}\right)^{-p}\\
    =&\frac{r}{q^p} \Big(1+p \frac{r-1}{q}+p(p+1)(\frac{r-1}{q})^2 +\cdots \Big).
\end{split}
\end{equation}
If the first two terms in this Taylor expansion are used for the high-order convolutions $\mathbf{J}^{*r_1}(q)$ and $ \mathbf{J}^{*r_2}(q)$ involved in (\ref{tay0}), then $\alpha_r$ can be expressed as follows for $0<r_1<q$ and $0<r_2<q$
\begin{equation}\label{cancel}
\begin{split}
    \alpha_r =\frac{1}{(2r)!q^{2p}}\sum_{r_1+r_2=2r} &\Bigg (\binom{2r}{r_1} (-1)^{r_1}\Big(r_1+\frac{pr_1(r_1-1)}{q}\Big) \\&\Big(r_2+\frac{pr_2(r_2-1)}{q}\Big) \Bigg )=0,
\end{split}
\end{equation}
due to the binomial theorem. 

\par  Equation (\ref{cancel}) shows that even though many hopping pathways contribute to transport entanglement through the chain, their effects cancel out due to accumulated phase differences, which as we will further demonstrate, ultimately enforces the emergence of light cones. Hence, equation (\ref{cancel}) manifests the destructive interference phenomenon. It demonstrates that despite the extensive accumulation of hopping amplitudes in the high-order convolutions $\mathbf{J}^{* r_1} (q)$ and $\mathbf{J}^{* r_2} (q)$, the main body of effects contributing to the buildup of $\alpha_r$ cancels out due to accumulated phases during the hopping process. This cancellation of entangling effects at distance $q$ leads to $\alpha_r=0$ up to the considered order of approximations (see Figure \ref{fig22}).

\par To derive (\ref{cancel}) and show that $\alpha_r = 0$, we truncated the expansions (\ref{expansionJt1}) and (\ref{tay}). Although such cancellations occur for other terms in both expansions, as demonstrated in the full proof \cite{suppl}, the cancellation is not complete for higher-order terms in (\ref{tay}) \cite{explanation4}. This results in a marginal contribution to $\alpha_r$ from these higher-order terms, which is {factorially small in $q$} and, for concreteness, is characterized in the proof \cite{suppl} of the theorem in the next subsection. The theorem in the next subsection concisely accounts for all effects beyond the truncations used in this subsection to show that (\ref{cancel}) ultimately leads to formation of the $2r<q$ light cone.

\par It is important to note that since (\ref{cancel}) vanishes regardless of the value of $p$, the early entanglement dynamics and the structure of the emergent light cone become increasingly independent of the parameter $p$ when $p>2$, explaining prior observations \cite{PhysRevLett.124.180601}. We will use this feature in subsection (\ref{nec}) to prove the necessity of destructive interference in explaining the formation of light cones.

\subsection{{Emergence of the $2r\approx q$ light cone due to destructive interference (sufficiency)}} \label{sectheorem}
\par In the previous subsection, we showed that the main body of entangling effects traveling to distance $q$ destructively interfere and cancel out when $r_1 < q$ and $r_2 < q$. To ensure the validity of these inequalities, and given that $2r = r_1 + r_2$, we consider the region $2r < q$. Therefore, for the $2r < q$ light cone, by taking into account all higher-order terms in the expansions (\ref{expansionJt1}) and (\ref{tay}), the following theorem is proved in Section III of \cite{suppl}.

\par\vspace{0.2cm} \textbf{Theorem:}\\ \textit{For spin chains governed by the long-range interacting Hamiltonian (\ref{hamil22})-(\ref{PD}), when $p>2$, the entanglement measure $\mathcal{Q}_{q}(t)$, defined in (\ref{tay0}), is constrained within linear light cone $2r<q$, as defined in Section \ref{lightchar}. Specifically, for $2r<q$ with $q\gg 1$, due to destructive interference among hopping terms, the coefficients $\alpha_r$ scale as follows }
\begin{equation}
    \alpha_r\sim {O}\bigg(\frac{p^4}{q^{2p+4}(2r-2)!} \bigg).
\end{equation}
\\See the proof in \cite{suppl}.

\par Notably, this scaling exhibits \textit{exponential} suppression in $q$ relative to the scaling of $\alpha_r$ required for an emergent light cone. We now elaborate.
Consider the case of NN interactions, where the first nonzero $\alpha_r$ corresponds to $r = q$, and in this case, we have $\alpha_r = \frac{1}{(r!)^2}$ \cite{azodi2023exact}. This leads to the formation of the $r=q$ light cone. Similarly, for the $2r \approx q$ light cone to emerge, the first nonzero $\alpha_r$ term (here $r \approx q/2$) must be of order $\hat{\alpha}_r = \frac{1}{(q/2)!^2}$. However, for $2r \approx q$, the scaling for $\alpha_r$ is given by:
\begin{equation}
    \alpha_r\sim {O}\bigg(\frac{p^4}{q^{2p+4}(q-2)!} \bigg),
\end{equation}
which is exponentially smaller than $\hat{\alpha}_r = \frac{1}{(q/2)!^2}$, as
\begin{equation}
    \frac{\alpha_r}{\hat{\alpha}_r}=\frac{p^4}{q^{2p+4}}\frac{(q/2)!^2}{(q-2)!}\sim \frac{1}{q^{2p+2} \binom{q}{q/2}} \sim \frac{1}{q^{2p+\frac{3}{2}}2^q }\rightarrow 0,
\end{equation}
{for } $q\gg 1$, where w have used the Stirling's approximation.


\begin{figure}
    \centering
    \begin{tikzpicture}[>=stealth]

   \begin{scope}[ xscale=0.85]
          \coordinate (dd) at (0,-0.2);

            \draw node[vertex1] (a) at (1,0) {};
   
  \draw node[vertex, color= darkred!95!darkblue] (b) at (2,0) {};
  \draw node[vertex, color= darkred!90!darkblue] (c) at (3,0) {};
  \draw node[vertex, color= darkred!85!darkblue] (d) at (4,0) {};
  \draw node[vertex, color= darkred!80!darkblue] (e) at (5,0) {};
  \draw node[vertex, color= darkred!75!darkblue] (f) at (6,0) {};
  \draw node[vertex, color= darkred!70!darkblue] (g) at (7,0) {};
  \draw node[vertex, color= darkred!65!darkblue] (h) at (8,0) {};
  \draw node[vertex, color= darkred!60!darkblue] (i) at (9,0) {};

  \draw[>->,, thick, darkred] (a) .. controls (2,-0.3) .. (c);
  \draw[>->,, thick, darkred] (c) .. controls (3.5,-0.3) .. (d);
  \draw[>->,, thick, darkred] (d) .. controls (5,-0.45) and (8,-0.45) .. (i);
    \draw[>->,, thick, darkred] (a) .. controls (1.5,0.3) .. (b);
  \draw[>->,, thick, darkred] (b) .. controls (2.5,0.3) .. (c);
  \draw[>->,, thick, darkred] (c) .. controls (4,0.45) and (8,0.45) .. (i);

  \draw[-, bluegrey]  (a) -- (b) (b)--(c) (c) --(d) (d)--(e) (e)--(f) (f)--(g) (g)--(h) (h)--(i);
  \draw[->, bluegrey, thick] (i)-- +(0.5,0);
    \draw[->, bluegrey, thick] (a)-- +(-0.5,0);
   \node at (3,0.5) {\scriptsize $\mathsf{r_1=3}$ }; \node at (3,-0.5) {\scriptsize  $\mathsf{r_2=3}$ };
\begin{scope}[yshift=-1.5cm]
           \draw node[vertex1] (a) at (1,0) {};
           \node at (0.1,0) {$+$};
   
   \draw node[vertex, color= darkred!90!darkblue2] (b) at (2,0) {};
  \draw node[vertex, color= darkred!83!darkblue2] (c) at (3,0) {};
  \draw node[vertex, color= darkred!75!darkblue2] (d) at (4,0) {};
  \draw node[vertex, color= darkred!65!darkblue2] (e) at (5,0) {};
  \draw node[vertex, color= darkred!55!darkblue2] (f) at (6,0) {};
  \draw node[vertex, color= darkred!45!darkblue2] (g) at (7,0) {};
  \draw node[vertex, color= darkred!35!darkblue2] (h) at (8,0) {};
  \draw node[vertex, color= darkred!25!darkblue2] (i) at (9,0) {};

  \draw[>->,, thick, darkred!100!black] (a) .. controls (2.4,-0.44) and (5.6,-0.44) .. (g);
  \draw[>->,, thick, darkred!100!black] (g) .. controls (8,-0.35) .. (i);    \draw[>->,, thick, darkred!100!black] (a) .. controls (2,0.4) and (5,0.4) .. (f);
  \draw[>->,, thick, darkred!100!black] (f) .. controls (6.5,0.3) .. (g);
  \draw[>->,, thick, darkred!100!black] (g) .. controls (7.5,0.3) .. (h);
  \draw[>->,, thick, darkred!100!black] (h) .. controls (8.5,0.3) .. (i);
  \node at (3,0.5) {\scriptsize $\mathsf{r_1=4}$ }; \node at (3,-0.5) {\scriptsize  $\mathsf{r_2=2}$ };
\end{scope}


  \draw[-, bluegrey]  (a) -- (b) (b)--(c) (c) --(d) (d)--(e) (e)--(f) (f)--(g) (g)--(h) (h)--(i);
  \draw[->, bluegrey, thick] (i)-- +(0.5,0);
    \draw[->, bluegrey, thick] (a)-- +(-0.5,0);

\begin{scope}[yshift=-3cm]
           \draw node[vertex1] (a) at (1,0) {};
                      \node at (0.1,0) {$+$};
   
  \draw node[vertex, color= darkred!93!darkblue3] (b) at (2,0) {};
  \draw node[vertex, color= darkred!86!darkblue3] (c) at (3,0) {};
  \draw node[vertex, color= darkred!76!darkblue3] (d) at (4,0) {};
  \draw node[vertex, color= darkred!65!darkblue3] (e) at (5,0) {};
  \draw node[vertex, color= darkred!55!darkblue3] (f) at (6,0) {};
  \draw node[vertex, color= darkred!48!darkblue3] (g) at (7,0) {};
  \draw node[vertex, color= darkred!40!darkblue3] (h) at (8,0) {};
  \draw node[vertex, color= darkred!36!darkblue3] (i) at (9,0) {};

  \draw[>->,, thick, darkred] (a) .. controls (2,0.5) and (7,0.5) .. (i);
  \draw[>->,, thick, darkred] (b) .. controls (3,-0.4) and (5,-0.4) .. (f);
    \draw[>->,, thick, darkred] (a) .. controls (1.5,-0.3)  .. (b);
  \draw[>->,, thick, darkred] (f) .. controls (6.5,-0.3) .. (g);
  \draw[>->,, thick, darkred] (g) .. controls (7.5,-0.3) .. (h);
  \draw[>->,, thick, darkred] (h) .. controls (8.5,-0.3) .. (i);
     \node at (3,0.5) {\scriptsize $\mathsf{r_1=1}$ }; \node at (3,-0.5) {\scriptsize  $\mathsf{r_2=5}$ };

    \draw[-, bluegrey]  (a) -- (b) (b)--(c) (c) --(d) (d)--(e) (e)--(f) (f)--(g) (g)--(h) (h)--(i);
  \draw[->, bluegrey, thick] (i)-- +(0.5,0);
    \draw[->, bluegrey, thick] (a)-- +(-0.5,0);
    \node at (4.5,-0.7) {$\vdots$};
    \draw [ thick, darkblue] (0,-1.2)--(9.5,-1.2);
\end{scope}
\begin{scope}[yshift=-4.7cm]
           \draw node[vertex1] (a) at (1,0) {};
      \node at (0.1,-0.03) {$=$};
   
  \draw node[vertex, color= darkred!95!darkblue] (b) at (2,0) {};
  \draw node[vertex, color= darkred!94!darkblue] (c) at (3,0) {};
  \draw node[vertex, color= darkred!92!darkblue] (d) at (4,0) {};
  \draw node[vertex, color= darkred!25!darkblue] (e) at (5,0) {};
  \draw node[vertex, color= darkred!10!darkblue] (f) at (6,0) {};
  \draw node[vertex, color= darkred!5!darkblue] (g) at (7,0) {};
  \draw node[vertex, color= darkred!5!darkblue] (h) at (8,0) {};
  \draw node[vertex, color= darkred!5!darkblue] (i) at (9,0) {};
  \node[below, yshift=-2pt] at (i) {\footnotesize$q$};
 \node[below, yshift=-2pt] at (b) {\footnotesize$1$};
  \node[below, yshift=-2pt] at (c) {\footnotesize$2$};
   \node[below, yshift=-2pt] at (d) {\footnotesize$3$};


    \draw[-, bluegrey]  (a) -- (b) (b)--(c) (c) --(d) (d)--(e) (e)--(f) (f)--(g) (g)--(h) (h)--(i);
  \draw[->, bluegrey, thick] (i)-- +(0.5,0);
    \draw[->, bluegrey, thick] (a)-- +(-0.5,0);

\end{scope}
    
\end{scope}
        
    \end{tikzpicture}
    \caption{Destructive interference as the mechanism behind light cone formation in long-range interacting spin chains. The arrows depict the long-range hopping process connecting the spins, as discussed in Figure \ref{fig11}. Each contributing term to entanglement production at distance $q$ (illustrated in the top three spin chains) consists of two sequences of hoppings, shown above and below the chains, corresponding to $\mathbf{J}^{*r_1}(q)$ and $\mathbf{J}^{*r_2}(q)$ in (\ref{tay0}). The order of hoppings ($r_1$ and $r_2$) in each example sums to $2r = r_1 + r_2 = 6$. These hoppings generate entanglement at distance $q$, represented by the intensity of the orange color in the spins at $q$, in contrast to dark blue (signifying no entanglement production). However, depending on $r_1$ and $r_2$, the accumulated entangling effects acquire a phase. Crucially, due to this phase accumulation, the total entanglement built up at distance $q$ (when $q > r = 3$)—depicted in the lowest spin chain—is exponentially suppressed as a result of destructive interference between different hopping pathways. This suppression gives rise to an emergent effective light cone, restricting entanglement propagation despite the long-range nature of interactions. The abrupt transition from orange to blue in the lowest spin chain visually represents this suppression. The proof in Section \ref{secIV} establishes that destructive interference is both necessary and sufficient for the formation of light cones, underscoring the fundamental role of this phenomenon in explaining the persistence of locality in long-range interacting spin chains.
}
  
    \label{fig22}
\end{figure}

\subsection{The necessity of destructive interference for emergent light cones}\label{nec}
The destructive interference mechanism is not only sufficient (as described in the previous section) but also necessary for the consistency of observed results in the literature. 

\par Prior studies have indicated a robust and universal emergence of \textit{linear} light cones in LRI systems when the range of interactions exceeds a threshold that varies depending on the nature of the system \cite{kuwahara2020strictly, tran2020hierarchy, PhysRevLett.124.180601, chatterjee2016multiple, tran2020hierarchy}. Here, the term {robust} signifies that the linearity of the light cone remains independent of \( p \) once it surpasses the threshold. 

On the other hand, in Section (\ref{lightchar}), we demonstrated that for linear light cones to emerge, it is necessary for \( \alpha_r \) to vanish for \( r < q \) (or \( 2r < q \), depending on the light cone). These two statements together dictate that the condition \( \alpha_r \approx 0 \) must remain independent of \( p \), meaning 
$\frac{d^s}{dp^s} \alpha_r = 0 \quad \text{for all } s = 1,2,\dots$.

Logically, such a condition can arise only under one of two scenarios:  
\begin{enumerate}  
    \item \( p \) is entirely absent from the formulation of entanglement—which is demonstrably not the case (see (\ref{tay0})), or  
    \item despite \( p \) appearing in the formulation, the value of \( \alpha_r \) remains fully independent of \( p \).  
\end{enumerate}  

Since \( p \) is explicitly present in the hopping processes—being embedded in the higher-order convolutions inside the summation that defines \( \alpha_r \)—its presence should, in principle, lead to a dependence of \( \alpha_r \) on \( p \). The only way this can be true is if the contributions of \( p \) within the hopping processes systematically cancel out, forcing \( \alpha_r \approx 0 \) for all \( p \).  

This can only occur through cancellation of terms (i.e., destructive interference) among the hopping processes (that involve $p$), as in (\ref{cancel}). Consequently, destructive interference is not just an incidental effect but a necessary mechanism ensuring the independence of the light cone structure from \( p \), thereby reinforcing the robustness of the emergent linearity in the emergent light cone structure.

\section{Faster entanglement propagation for shorter-ranged interactions}
\par The destructive interference phenomenon has a counter-intuitive implication. In particular, the entanglement edge must propagate faster when the interaction function is shortened, specifically when the interaction terms that extend beyond the range ($\eta$) are {not present in} the interaction function. This phenomenon is the direct consequence of destructive interference in the early entanglement region due to the following reason. The destructive interference occurs when \textit{all} entangling effects are superposed (see (\ref{cancel})). Therefore, if part of the stream of quasi-particles is obstructed due to shortening the interaction function, then the destructive interference is weakened. Hence, the entanglement edge propagates faster. This phenomenon is mathematically proved in Section V of \cite{suppl}, and numerically demonstrated in the next subsection. Here, a shortened interaction refers to:
    \begin{equation}\label{sharp}
        \mathbf{J}_{\eta}(n)=
\begin{cases}
  \frac{J}{|n|^p}, & 0<|n|\leq \eta\\
  0, & |n|>\eta, n=0,
\end{cases}
    \end{equation}
where $\eta=1$ gives {NN} interaction.
\subsection{Numerical Evidence}
 Figure \ref{fig:delayed} shows the Entanglement Edge Times (EETs) at a distance of $q=10^3$ from the quenched spin in a ferromagnetic Heisenberg chain (when $H_{j,k}= \mathbf{S}_j \mathbf{S}_k$, with $\mathbf{S}_j$  being the spin operator at site $j$) as a function of the interaction range ($\eta$) and for different values of $p$. EET represents the time required for entangling effects, beyond quantum tunneling, to reach spins at the given distance. {In this analysis, EET denotes the time when the entanglement measure $\mathcal{Q}_q(t)$ exceeded $10^{-5}$.} Prolonged EETs for the full-range interaction (denoted by $\eta=\infty$), compared to shortened interaction ranges demonstrate the destructive interference effect, which is apparent for $p\geq2$ among the simulated cases. In Figure \ref{fig:delayed}, the orange dashed lines show prolonged EETs when $\eta=\infty$ is contrasted with $\eta=14$. Destructive interference also manifests itself in the upward slopes of the curves, highlighted in orange, which indicate an increase in the EETs as the interaction range grows. 
\par Another notable aspect in Figure \ref{fig:delayed} is the lack of dependence of the $\eta=\infty$-EET on the variable $p$ for $p> 3$. This can be seen in the flattening of the curves as $p$ increases, which is caused by destructive interference. To illustrate this phenomenon more clearly, the inset in figure \ref{fig:delayed} displays the EETs for the full-range interaction length as a function of $p$. Consequently, for $p > 3$, as $\eta$ increases and destructive interference appears, the EETs quickly revert to the NN interaction time ($\eta=1$), indicated by the horizontal dashed black line. Thus, in this region, the EETs are not only independent of $p$, but also become increasingly independent of $\eta$ when $\eta$ exceeds a certain value. Based on this feature, and given that $p\rightarrow \infty$ corresponds to NN interactions, for $p>3$, light cones must be strictly linear (similar to the case of NN interactions), confirming the results in \cite{kuwahara2020strictly}.

\subsection{The Practical Softening of Interactions}
\par {As an alternative to the sharp shortening (\ref{sharp}), which can be practically challenging to implement, the experimentally feasible \cite{yanay2022mediated, hollerith2022realizing} softened shortening (characterized by the rate $\sigma$) is given by}
   \begin{equation}\label{softss}
        \mathbf{J}_{\eta,\sigma}(n)=
\begin{cases}
  \frac{J}{|n|^p}, &0< |n|\leq \eta \\
  \frac{J}{|n|^p} e^{-\sigma (|n|-\eta)}, & |n|>\eta
\end{cases}
    \end{equation}
This approach can also be used to experimentally observe the counterintuitive acceleration of entanglement transport in the chain. Table \ref{TT1} shows the EET values for various softened interactions in a numerical simulation of ferromagnetic Heisenberg chains, described in the previous subsection. In this table, $\eta = 5$ represents the interaction range for shortened interactions, while $\eta = \infty$ denotes full long-range interaction. Additionally, $\sigma = \infty$, 1.5, and 0.5 correspond to sharp, semi-softened, and highly softened shortening of the interaction function, respectively.

\par As the simulation data clearly indicate, the EETs for shortened interactions are distinctly lower compared to the full long-range case ($\eta = \infty$). This difference becomes more pronounced with increasing sharpness of shortening. Therefore, even though sharp shortening might not be feasible in experimental settings, we propose that softened shortening can effectively be used to demonstrate the destructive interference phenomenon in the laboratory.

\begin{table}[h!]

\centering
\begin{tabular}{ |p{3cm}||p{1.2cm}|p{1.2cm}|p{1.2cm}|p{1.2cm}|  }
 \hline
 \multicolumn{5}{|c|}{EETs for softened interaction shortening} \\
 \hline
& $\eta=5$ $\sigma=\infty$ &$\eta=5$ $\sigma=1.5$& $\eta=5$ $\sigma=0.5$&$\eta=\infty$\\
 \hline
{{EET ($\frac{\hbar}{J}$)} ($p=2.5)$} & 413.0  &   417.0 &431.6&455.9  \\
 \hline
 
\end{tabular}
\caption{Entanglement Edge Time (EET) for softened interaction functions. Parameters $\eta$ and $\sigma$ represent the interaction range and shortening sharpness, respectively (see \ref{softss}). $\eta=\infty$ denotes full long-range interaction, and $\sigma=\infty$ denotes sharp shortening. EETs are calculated for $p=2.5$ and are given in units of $\frac{\hbar}{J}(\pm 0.1)$. The data in the table clearly demonstrate that even with soft shortening, which is expected to be experimentally feasible in an appropriate laboratory setting, the counterintuitive acceleration of entanglement persists.  }
\label{TT1}
\end{table}

\usetikzlibrary{shapes}

\begin{figure}
    \centering
        \begin{tikzpicture}[]

        \begin{axis}[xmin=1,
        xmax=21,
        ymin=0,
        ymax=550,
        xtick={2,...,15,17},
        xticklabels={1,,3,,5,,7,,9,,11,,13,,$\infty$},
        ylabel={Entanglement Edge Time (EET) $(\frac{\hbar}{J})$},
        xlabel={Interaction range-$\eta$},
        width = 0.5\textwidth,
    	height = 0.57\textwidth,
        legend pos=north west]
    
  \addplot [ color=darkblue, mark=o]    table [x=range, y=e17]{100-2501-edge.txt};
  \addplot [ mark=o, darkblue!90!black,dashed,mark options={solid,fill=darkblue!90!black}]    table [x=range, y=e17]{1000-2501-infty.txt};

  \node at (19,110) {$p=1.7$};
    \addplot [ mark=o, darkblue!90!black]    table [x=range, y=e19]{100-2501-edge.txt};
  \addplot [ mark=o, darkblue!90!black,dashed,mark options={solid,fill=darkblue!90!black}]    table [x=range, y=e19]{1000-2501-infty.txt};
   \node at (19,207) {$p=1.9$};

    \addplot [ mark=o, darkblue!90!black]    table [x=range, y=e20]{100-2501-edge.txt};
  \addplot [ mark=o, darkred!80!black,dashed,mark options={solid,fill=darkred!80!black}]    table [x=range, y=e20]{1000-2501-infty.txt};

  \node at (19,303) {$p=2$};
      \addplot [ mark=o, darkblue!90!black]    table [x=range, y=e21]{100-2501-edge.txt};
  \addplot [ mark=o, darkred!80!black,dashed,mark options={solid,fill=darkblue!90!black}]    table [x=range, y=e21]{1000-2501-infty.txt};
     \node at (19,375) {$p=2.1$};
        \addplot [ mark=o, darkblue!90!black]    table [x=range, y=e23]{100-2501-edge.txt};
  \addplot [ mark=o, darkred!80!black,dashed,mark options={solid,fill=darkblue!90!black}]    table [x=range, y=e23]{1000-2501-infty.txt};
       \node at (19,429) {$p=2.3$};
          \addplot [ mark=o, darkblue!90!black]    table [x=range, y=e25]{100-2501-edge.txt};
  \addplot [ mark=o, darkred!80!black,dashed,mark options={solid,fill=darkblue!90!black}]    table [x=range, y=e25]{1000-2501-infty.txt};
         \node at (19,456) {$p=2.5$};
            \addplot [ mark=o, darkblue!90!black]    table [x=range, y=e29]{100-2501-edge.txt};
  \addplot [ mark=o, darkred!80!black,dashed,mark options={solid,fill=darkred!80!black}]    table [x=range, y=e29]{1000-2501-infty.txt};
             \node at (19,479) {$p=2.9$};
\addplot [ mark=o, darkblue!90!black]    table [x=range, y=e35]{100-2501-edge.txt};
  \addplot [ mark=o, darkred,dashed,mark options={solid,fill=darkred}]    table [x=range, y=e35]{1000-2501-infty.txt};
             \node at (19,500) {$p=3.5$};


    \addplot [ thick,darkred]    table {
5 373
6 373
7 375
8 378
9 383
10 387
11 391
12 396
13 401
14 404
15 407
    
    };

        \addplot [ thick,darkred]    table {
5 406
6 413
7 423
8 432
9 443
10 447
    };

    \addplot [ thick,darkred]    table {
3 441    
4 444  
5 459
6 476
7 493

    };

        \addplot [ thick,darkred]    table {
3 465
4 479  
5 494
6 499

    };


    \draw [ultra thin, black ,dashed ] (0,491) -- (18,491);



          \end{axis}

         \draw [color=white,  fill=white] (5.53, 0) circle [radius=0.05];
        \node  at (5.45,0) {\scriptsize $\backslash$};
        \node  at (5.58,0) {\scriptsize $\backslash$};

                 \draw [color=white,  fill=white] (5.53, 8.64) circle [radius=0.05];
        \node  at (5.45,8.64) {\scriptsize $\backslash$};
        \node  at (5.58,8.64) {\scriptsize $\backslash$};

\begin{scope}[xshift=1.4cm, yshift=0.85cm]

    \begin{axis}[xmin=1,
        xmax=7,
        ymin=280,
        ymax=550,
        ytick={300,400,500},
        xtick={2,...,5,6.5},
        xticklabels={2,...,5,$\infty$},
        xlabel={$p$},
        ylabel={ {EET} ($\eta=\infty$)},
        width = 0.27\textwidth,
    	height = 0.19\textwidth,
        legend pos=north west,
        every axis x label/.style={yshift=-5mm,right=15mm}
        ]


    \addplot [ only marks,mark=o, darkred,mark options={solid,fill=red!80!black}]    table [x=p, y=time]{edge-for-long-range.txt};
    \draw [ black ,dashed ] (0,491) -- (11,491);
    
    \end{axis}
        \draw [color=white,  fill=white] (2.55, 0) circle [radius=0.05];
        \node  at (2.48,0) {\scriptsize $\backslash$};
        \node  at (2.58,0) {\scriptsize $\backslash$};

     \draw [color=white,  fill=white] (2.55, 1.82) circle [radius=0.05];
        \node  at (2.48,1.82) {\scriptsize $\backslash$};
        \node  at (2.58,1.82) {\scriptsize $\backslash$};
\end{scope}

\end{tikzpicture}
    \caption{Entanglement Edge Times (EETs) for various interaction ranges ($\eta$) and different values of $p$. The plot demonstrates how shortening interactions alters entanglement propagation. The orange-highlighted upward slopes of the curves illustrate the unexpected increase in EETs as the interaction range ($\eta$) grows for $p> 2$. Furthermore, the prolonged EETs for full-range interaction (represented by $\eta=\infty$) compared to shortened interaction values are shown by orange dashed lines. Both of these unusual phenomena result from the destructive interference between entangling effects, which leads to the formation of an effective light cone in long-range interacting spin chains. The inset shows the progress of the $\eta=\infty$-EETs versus $p$. This figures shows the quick convergence of EETs to the nearest-neighbor value for $p> 3$.  }
    \label{fig:delayed}
\end{figure}

\section{Concluding remarks} 
\par 

In this work, 
we prove that quantum destructive interference, a physical mechanism that, to the best of our knowledge, has not been identified in the literature, as both necessary and sufficient for the emergence of light cones in quantum spin chains with long-range interactions. As a result, we identify a previously unrecognized inherent insulating behavior that restricts entanglement propagation beyond the light cone. Specifically, we demonstrate that while entangling effects propagate rapidly due to the long-range nature of interactions, they are significantly suppressed by the destructive interference mechanism. 

\par The results of this paper not only provide new theoretical insights into the notion of locality (as a result of the emergent light cones) in quantum systems but also offer a concrete example of how causal structures, such as light cones, emerge within the quantum dynamics of many-body systems. This raises intriguing questions about the role of destructive interference in embedding relativistic principles within quantum many-body dynamics. Whether relativity itself is an emergent phenomenon remains one of the most fundamental questions in theoretical physics \cite{carlip2014challenges}, explored in contexts such as holographic spacetime \cite{maldacena1999large, ryu2006holographic}, and other frameworks \cite{verlinde2017emergent, jacobson1995thermodynamics, doi:10.1142/S0218271810018529}. Our work introduces destructive interference as a candidate mechanism that could bridge quantum dynamics and emergent causal structures, a direction for further investigation. 

\par This analysis also unifies prior findings in the literature, including the emergence of Lieb-Robinson bounds in power-law decaying interactions with $p>2$ \cite{tran2020hierarchy}, which coincides with the breakdown of destructive interference. Furthermore, the presence of a strictly linear light cone for $p>3$ \cite{kuwahara2020strictly, chen2019finite} aligns with our result that, in this regime, destructive interference constrains entanglement propagation to mimic that of nearest-neighbor interactions. 

\par We predict that the destructive interference phenomenon, underlying the counterintuitive acceleration of entanglement growth, can be experimentally observed through the use of shortened interactions. While exact (sharp) shortening may pose experimental challenges, softened shortening yield similar acceleration effects. Notably, similar tunable shortening have already been implemented in Rydberg atom arrays \cite{bernien2017probing}, making them a promising platform for testing our predictions. Additional platforms for investigation include cold atoms \cite{cheneau2012light, rigol2008thermalization} and trapped ions \cite{jurcevic2014quasiparticle, katz2023demonstration, britton2012engineered, feng2023continuous}. Future work should investigate whether destructive interference persists in the strong long-range interacting limit $p<1$, potentially uncovering new mechanisms for entanglement suppression.

\begin{acknowledgments}
P.A would like to thank Hans Halvorson for fruitful discussions. P.A acknowledges partial support from the Princeton Program in Plasma Science and Technology (PPST). H.R and P.A acknowledge partial support from the U.S Department Of Energy (DOE) grant (DE-FG02-02ER15344).
\end{acknowledgments}

\bibliography{citations}

\begin{thebibliography}{61}
\bibitem [{exp({\natexlab{a}})}]{explanation5}%
  \BibitemOpen
  \href@noop {} {\bibinfo {title} {For brevity, in the remainder of the paper, we will avoid using the phrase effective light cones, and will simply use light cones.}} ({\natexlab{a}})\BibitemShut {NoStop}%
\bibitem [{exp({\natexlab{b}})}]{explanation3}%
  \BibitemOpen
  \href@noop {} {} ({\natexlab{b}}),\ \bibinfo {note} {this conclusion is valid becasue $\alpha_q=\frac{1}{(q! ^2)}$. Upon using the Stirling's approximation, we get (\ref{first}). Higher order terms, proportional to $(\frac{t}{q+1})^{2(q+1)}, (\frac{t}{q+2})^{2(q+2)}$, ect. follow similarly. Importantly, given the structure of these terms, for $t<q$, the first term in (\ref{first}) is exponentially dominant.}\BibitemShut {Stop}%
\bibitem [{exp({\natexlab{c}})}]{explanation1}%
  \BibitemOpen
  \href@noop {} {} ({\natexlab{c}}),\ \bibinfo {note} {this identification is valid because if higher-order terms, e.g., $r=q+1$, $r=q+2$, etc., were to dominate the series for arbitrary time $t<q$, then the coefiicients of these terms would need to grow exponentially with $r$ to compensate for the exponentially smaller parts $(\frac{t}{q+1})^{2(q+1)}$, etc.}\BibitemShut {Stop}%
\bibitem [{exp({\natexlab{d}})}]{explanation4}%
  \BibitemOpen
  \href@noop {} {} ({\natexlab{d}}),\ \bibinfo {note} {when the term of order $s$ in (\ref{tay}), i.e., corresponding to $(\frac{r_1-1}{q})^s$ is substituted for $\mathbf{J}^{*r_1}(q)$ in (\ref{tay0}), all of the terms with $r_1\geq s$ cancel out in the summation. Therefore, only the first $s-1$ terms remain, which are exponentially (in $s$) suppressed outside the light-cone $r<q$}\BibitemShut {NoStop}%
\end{thebibliography}

\end{document}